\documentclass[iop]{emulateapj}
\usepackage{color,psfig, epsfig}

\newcommand{\swift}{{\em Swift}}
\newcommand{\chandra}{{\em Chandra}}

\newcommand{\myemail}{roberto.soria@curtin.edu.au}

\slugcomment{To appear in ApJ}

\shorttitle{Slim-disk ULX in M\,83}
\shortauthors{Soria et al.}

\begin{document}


\title{The slim-disk state of the ultraluminous X-ray source in M\,83}


\author{Roberto Soria}
\affil{International Centre for Radio Astronomy Research, Curtin University, GPO Box U1987, Perth, WA 6845, Australia}
\email{\myemail}

\author{K. D. Kuntz}
\affil{The Henry A. Rowland Department of Physics and Astronomy, Johns Hopkins University, 3400 N. Charles Street, Baltimore, MD 21218, USA}

\author{Knox S. Long}
\affil{Space Telescope Science Institute, 3700 San Martin Drive, Baltimore, MD 21218, USA}

\author{William P. Blair}
\affil{The Henry A. Rowland Department of Physics and Astronomy, Johns Hopkins University, 3400 N. Charles Street, Baltimore, MD 21218, USA}

\author{Paul P. Plucinsky}
\affil{Harvard-Smithsonian Center for Astrophysics, 60 Garden Street, Cambridge, MA 02138, USA}

\and

\author{P. Frank Winkler}
\affil{Department of Physics, Middlebury College, Middlebury, VT 05753, USA}








\begin{abstract}
The transient ULX in M\,83 that went into outburst in 
or shortly before 2010 is still active. Our new {\it XMM-Newton} spectra 
show that it has a curved spectrum typical of the upper end of 
the high/soft state or slim-disk state. It appears to be spanning the gap 
between Galactic stellar-mass black holes and the ultraluminous state, 
at X-ray luminosities $\approx 1$--$3 \times 10^{39}$ erg s$^{-1}$ 
(a factor of two lower than in the 2010--2011 {\it Chandra} observations).
From its broadened disk-like spectral shape at that luminosity, 
and from the fitted inner-disk radius and temperature, we argue that 
the accreting object 
is an ordinary stellar-mass black hole with $M \sim$$10$--$20 M_{\odot}$.
We suggest that in the 2010--2011 {\it Chandra} observations, 
the source was seen at a higher accretion rate, resulting  
in a power-law-dominated spectrum with a soft excess at large radii. 
\end{abstract}





\section{Introduction}
The most luminous sub-class of X-ray binaries is known as ultraluminious 
X-ray sources (ULXs). They are a heterogenous class of objects, 
empirically defined as non-nuclear 
accreting systems with an X-ray luminosity 
$L_{\rm X} \ga 3 \times 10^{39}$ erg s$^{-1}$ 
(\citealt{feng11} for a review). As such, ULXs are more luminous than ordinary 
stellar-mass black holes (BHs) in our own Galaxy, which typically 
peak at $L_{\rm X} \la 10^{39}$ erg s$^{-1}$.
The two main competing scenarios invoked to explain the bulk 
of the ULX population are higher BH masses and super-critical accretion.
The mass of a BH determines its critical accretion rate 
and the corresponding Eddington luminosity. Galactic BHs have masses 
$\approx$$10M_{\odot}$ \citep{kreidberg12} and a corresponding 
Eddington luminosity $L_{\rm Edd} \approx$$10^{39}$ erg s$^{-1}$; however,
stellar evolution models predict that BHs with masses 
up to $\approx 80 M_{\odot}$ \citep{belczynski10} may be formed via direct 
collapse of metal-poor stars of initial mass $> 120 M_{\odot}$. 
Alternatively, if the accretion rate is super-critical 
($\dot{m} \equiv 0.1 \dot{M}c^2 / L_{\rm Edd} > 1$), 
the accretion luminosity may exceed 
the Eddington limit slightly: 
$L \approx L_{\rm Edd} (1 + \ln \dot{m})$ for advection-dominated inflows, or 
$L \approx L_{\rm Edd} (1 + \frac{3}{5}\ln \dot{m})$ when accretion 
is limited by radiatively-driven outflows \citep{poutanen07,ss73}.
Testing between the two alternative scenarios (more massive BHs 
or super-Eddington accretion) has proven challenging, 
owing to the scarcity of kinematic mass measurements for BHs in ULXs.

There are indirect methods for estimating BH masses in X-ray binaries, 
based on their spectral state behaviour \citep{mcclintock06}.
At low or moderate accretion rates (low/hard state: 
$\dot{m} \la$ a few percent), 
BH X-ray binaries are dominated by a hard power-law spectrum 
produced in a hot, geometrically thick, optically thin 
Comptonizing region; they are also radio-loud, that is they sustain  
a steady relativistic jet.
At higher accretion rates (high/soft state: $\dot{m} \la 0.5$)
they are dominated by thermal emission from a geometrically thin, 
optically thick accretion disk, and the jet is quenched.
Low/hard and high/soft state are easily distinguished 
with X-ray and radio observations. The evolutionary track between 
the two states is also well modelled and fairly standard for all 
BH transients \citep{fender04}. Thus, the empirical identification 
of either state provides an order-of magnitude estimate of the Eddington 
ratio and therefore of the BH mass. 
Moreover, when a BH X-ray binary is in the high/soft state, 
a more robust estimate of the BH mass comes from the fitted inner radius 
and peak temperature of the accretion disk, 
based on standard disk models and on the assumption that the disk 
extends down to the innermost stable circular orbit \citep{kubota98,ss73}.
For even higher accretion rates ($\dot{m} \ga 0.5$), there is still 
no consensus 
about any direct relation between Eddington ratio and X-ray appearance. 
In the same range of observed luminosities, 
some ULXs appear power-law-dominated, others have a curved or thermal 
spectrum \citep{sutton13,soria11}. There is also no consensus about 
the underlying physical evolution of the accretion flow when the accretion rate 
reaches Eddington. In one scenario, the inflow can still be described 
as a (non-standard) accretion disk (slim disk models: \citealt{watarai01}).
Alternatively, many ULX X-ray spectra have been modelled 
with Comptonized emission from a warm ($kT \la 2$ keV), optically thick 
medium (corona and/or dense outflow) which covers or replaces 
the inner disk \citep{gladstone09,done06}.
Finally, it is still not clear whether power-law-like and curved ULX spectra 
correspond to different ranges of super-Eddington accretion rates, 
or instead to different viewing angles and outflow thickness 
\citep{sutton13,kawashima12}.
In this context, discovering and monitoring transient ULXs is crucial 
to disentangle intrinsic changes in the inflow and emission properties 
from orientation effects.

\subsection{Transient ULX in M\,83}
In 2010, we detected a new ULX  \citep{soria12} in an inter-arm region 
of the spiral galaxy M\,83 [$d = (4.6 \pm 0.2)$ Mpc: \citealt{saha06}]. 
Located at R.A.(2000) $= 13^{\rm h}37^{\rm m}05^{\rm s}.13 \pm 0^{\rm s}.01$, 
Dec.(2000) $= -29^{\circ}52'07\farcs{1} \pm 0\farcs{2}$, it is listed 
as source X299 in our full {\it Chandra} catalog of M\,83 sources 
\citep{long14}. Henceforth, we will refer to this source 
as M\,83 ULX-1, for simplicity. 
Finding a ULX in M\,83 is remarkable in itself, given the unusually high metal 
abundance (up to twice the solar abundance) of the inner disk in this galaxy 
\citep{bresolin02,pilyugin06,pilyugin10}. 
Solar-metallicity stars are not expected to produce BHs more massive 
than $\approx 15 M_{\odot}$ \citep{belczynski10}. Thus, the detection 
of M\,83 ULX-1 is already a strong argument in support of the interpretation 
of ULXs as super-Eddington sources rather than particularly heavy 
stellar BHs.

Although its luminosity is not extreme 
for a ULX ($L_{\rm X} \approx 5 \times 10^{39}$ erg s$^{-1}$ 
in the {\it Chandra} observations of 2011 March, with 
a plausible peak at $L_{\rm X} \approx 7 \times 10^{39}$ erg s$^{-1}$ 
in the {\it Swift} observations of 2011 February), 
M\,83 ULX-1 has the unusual property of being a transient. 
Most ULXs in nearby galaxies are variable (by a factor of a few) 
but remain persistently above $\sim$$10^{39}$ erg s$^{-1}$. 
For example, almost all ULXs found by {\it Einstein} in the 1980s and 
{\it ROSAT} in the 1990s are still bright today. Instead, 
ULX-1 was undetected prior to 2010, with an upper limit 
$L_{\rm X} \la 2 \times 10^{37}$ erg s$^{-1}$ in {\it ROSAT} 
(1993 observation) and {\it XMM-Newton} (2003 observation), 
and $L_{\rm X} \la 10^{36}$ erg s$^{-1}$ in a 2000 {\it Chandra} observation
\citep{soria12}.
In Galactic BH X-ray binaries, transient systems usually contain 
low-mass evolved donors, while persistent systems have 
more massive Be stars or supergiants. We showed \citep{soria12} 
that the donor of ULX-1 is indeed a relatively old star, 
probably a red giant with a mass $< 4M_{\odot}$ and an age $\ga$ a few 100 Myr.
The blue optical counterpart seen only during the outburst 
arises from a large X-ray-irradiated disk.

Because it is known to be a transient, it is plausible 
that ULX-1 will undergo state transitions analogous 
to those of transient stellar-mass BHs. If so, we can estimate its BH mass 
in two ways: a) by determining the luminosities at which the system transitions 
from the ultraluminous state to the canonical thermal dominant (high/soft) 
state, and then from the thermal dominant state to the low/hard state; 
and b) by fitting the disk parameters (temperature and innermost radius) 
while the source is in the thermal dominant state. 
Therefore, we have continued to monitor this source in an attempt 
to understand the intrinsic nature of ULXs and the connection 
between ULXs and Galactic BH X-ray binaries.


\begin{figure}
\hspace{-0.4cm}
\includegraphics[angle=0, scale=.73]{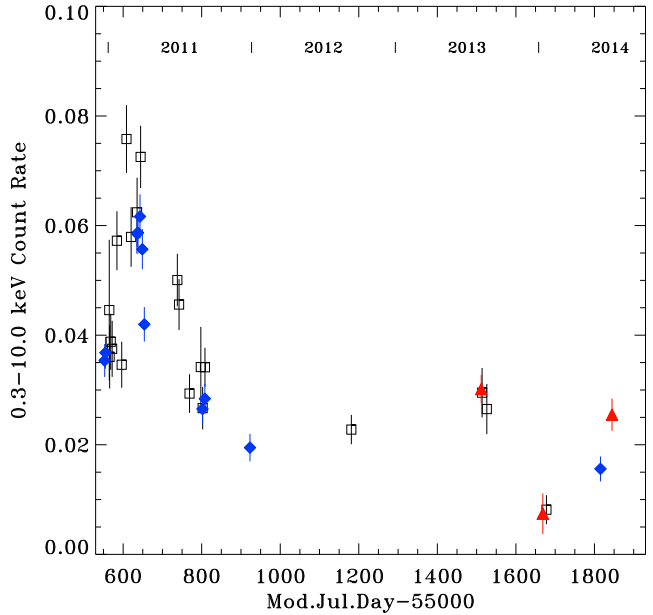}
\caption{X-ray lightcurve from all our observations 
between 2010 December and 2014 July (Table 1), expressed as  
an equivalent {\it Swift}/XRT count rate. 
The flux in the {\it Chandra}/ACIS and 
{\it XMM-Newton}/EPIC observations was converted to a {\it Swift}/XRT count 
rate by convolving the best-fitting spectral models 
with an XRT response matrix. Black squares are the {\it Swift}  
datapoints; blue lozenges are those from {\it Chandra}; red triangles 
those from {\it XMM-Newton}. 
\label{fig1}}
\vspace{0.4cm}
\end{figure} 


\section{New X-ray observations of M\,83 ULX-1}
Since our extensive multiband 2010--2011 campaign \citep{long14}, 
we have carried out a number of X-ray observations of M\,83 (Table 1).  
Specifically, we used the {\it Swift} X-Ray Telescope (XRT) to carry out 
short ($\approx$2--5 ks) observations 
on 2012 September 11, 2013 August 9 and 21, and 2014 January 20.
In addition, we used the EPIC cameras on board {\it XMM-Newton} to observe 
the ULX on 2013 August 7 ($\approx$50 ks), 2014 January 11 ($\approx$40 ks) 
and 2014 July 6 ($\approx$30 ks).
Finally, in the analysis reported here, we make use of a 30-ks 
{\it Chandra}/ACIS-I observation of M\,83 which took place on 2014 June 7, 
which PI Ann Hornschemeier has kindly made available to us.

\begin{figure}[t]
\hspace{-0.45cm}
\includegraphics[angle=0,scale=0.71]{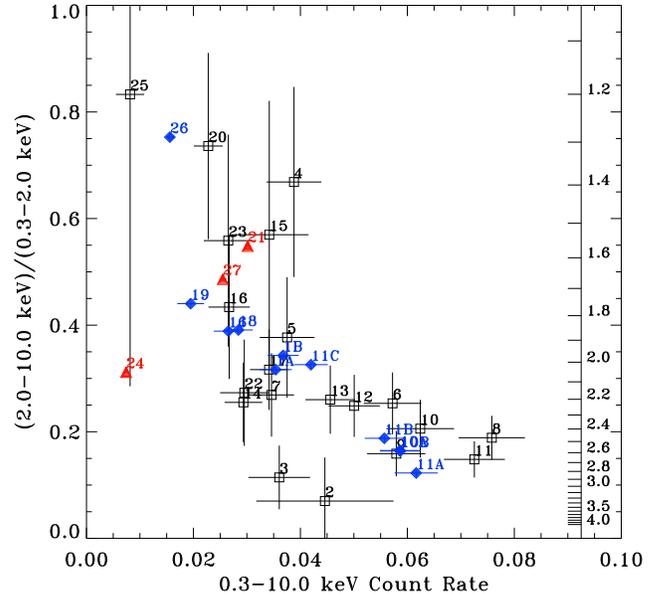}
\caption{Hardness ratio as a function of {\it Swift}/XRT count rate.
All hardness ratios have been converted to an equivalent {\it Swift}/XRT 
ratio of $(2$--$10)$ keV$/(0.3$--$2)$ keV count rates.
The equivalent single-component power-law index is shown on the right.
As in Figure 1, black squares are the {\it Swift}  
datapoints; blue lozenges are those from {\it Chandra}; red triangles 
those from {\it XMM-Newton}. Numbers refer to the sequential order 
of the observations (Table 1).
\label{fig2}}
\vspace{0.3cm}
\end{figure}

The {\it Swift} light curve (Figure~\ref{fig1}) was produced from the standard level-2 data products.
In all cases, the ULX was placed at the aim point, so that the response and the point spread function are well characterized.
The count rates in the various bands were extracted as described by \citet{soria12}.
Because the {\it Swift} monitoring data had rather short exposures, the number of counts is insufficient for spectral fitting, and determining the absolute flux is problematic. 
To avoid these issues, the light curve is expressed in {\it Swift}/XRT 
$0.3$--$10.0$ keV count rates.
We then placed observations from other missions (with much higher 
total counts per observation) onto this scale by fitting those spectra 
and applying the latest {\it Swift}/XRT response to the best-fit models.
The hardness ratio (Figure~\ref{fig2}) 
was expressed as the ratio of {\it Swift}/XRT count rates 
in the $2.0$--$10.0$ keV band over that in the $0.3$--$2.0$ keV band, 
and the measurements from the other missions were similarly placed 
on the {\it Swift}/XRT scale.
Since \citet{soria12}, the {\it Swift}/XRT responses appropriate for 2011 
have been changed significantly and the light curve has been updated 
accordingly. However, the overall response is not thought to have changed 
drastically in the 2011--2014 period over which these data were taken.

We processed the {\it XMM-Newton}/EPIC Observation Data Files with the 
Science Analysis System ({\small{SAS}}) version 13.0.0 (xmmsas\_201300501). 
The particle background was low during all three observations, 
so we did not need to filter out any exposure interval. 
We defined a circular source extraction region centered on the ULX, 
with a 20\arcsec radius, sufficiently small to avoid significant 
contamination from the nuclear starburst emission. 
We extracted the background from a composite region three times as large, 
suitably selected at the same distance from the galactic nucleus, 
and not including any other bright sources or chip gaps.
We selected single and double events 
(pattern $\la$4 for the pn and pattern $\la$12 for MOS1 and MOS2), 
with the standard flagging criteria \#XMMEA\_EP and \#XMMEA\_EM 
for the pn and MOS, respectively. After building response and ancillary 
response files with the {\small{SAS}} tasks {\it{rmfgen}} and {\it{arfgen}}, 
we used {\it{epicspeccombine}} to create average EPIC spectra 
and response files for each of the three epochs.
Finally, we grouped the spectra to a minimum of 20 counts per bin, 
so that we could use Gaussian statistics. 
We used {\small{XSPEC}} version 12.6 for spectral fitting \citep{arnaud96}.
For each epoch, we fitted the combined EPIC spectra, and we also fitted 
the three EPIC cameras simultaneously 
(leaving a free normalization factor between pn and MOS).
The results from the two methods are always consistent 
within the 90\% confidence limit, and the relative normalization 
factor between MOS and pn is $\approx 0.98$. 
In Table 2, we report the fit parameters for the combined EPIC spectra.  
For timing analysis, we used standard {\small{FTOOLS}} tasks 
\citep{blackburn95}.
Unlike most of the {\it Chandra}/ACIS-S spectra from 2010--2011, 
the {\it XMM-Newton}/EPIC spectra are not piled up 
(as shown by the {\small{SAS}} task {\it{epatplot}}).

We filtered and analyzed the {\it Chandra}/ACIS-I data with 
standard imaging and spectroscopic tasks  
({\it e.g.}, {\it dmcopy}, {\it dmextract}, and {\it specextract}), 
in the {\small{CIAO}} Version 4.6 \citep{fruscione06} data analysis system.
We used {\it PIMMS} Version 4.6b from the {\it Chandra} proposal 
planning toolkit to estimate the level of pile-up given the observed 
count rate, and found that it is $\approx$7\%.

\begin{deluxetable*}{lccccccc}
\tablecolumns{7}
\tabletypesize{\scriptsize}
\tablecaption{M83 X-ray Observations
\label{tab:obsid}}
\tablewidth{0pt}
\tablehead{
\colhead{Epoch} &
\colhead{Obsid} &
\colhead{Instrument} &
\colhead{Date} &
\colhead{Exposure} &
\colhead{Flux\tablenotemark{a}} &
\colhead{$L_X$\tablenotemark{b}} \\
\colhead{} &
\colhead{} &
\colhead{} &
\colhead{} &
\colhead{(s)} &
\colhead{($10^{-12}$ erg cm$^{-2}$ s$^{-1}$)} &
\colhead{($10^{39}$ erg s$^{-1}$)} }
\startdata
\cutinhead{{\it Swift}}
2  & 0031905002 & XRT & 2011-01-03 &  399 & $1.1\pm0.3$ & \\
3  & 0031905003 & XRT & 2011-01-04 & 1620 & $1.0\pm0.2$ & \\
4  & 0031905004 & XRT & 2011-01-07 & 2213 & $1.9\pm0.3$ &  \\
5  & 0031905005 & XRT & 2011-01-11 & 2140 & $1.4\pm0.2$ &  \\
6  & 0031905006 & XRT & 2011-01-23 & 2896 & $1.8\pm0.2$ &  \\
7  & 0031905007 & XRT & 2011-02-04 & 2938 & $1.1\pm0.1$ &  \\
8  & 0031905008 & XRT & 2011-02-16 & 2882 & $2.2\pm0.2$ &  \\
9  & 0031905009 & XRT & 2011-02-28 & 2863 & $1.6\pm0.2$ &  \\
10\tablenotemark{c} & 0031905010 & XRT & 2011-03-15 & 2285 & $1.9\pm0.2$ & \\
11\tablenotemark{d} & 0031905011 & XRT & 2011-03-24 & 3258 & $2.0\pm0.2$ &  \\
12 & 0031905012 & XRT & 2011-06-25 & 3240 & $1.6\pm0.2$ &  \\
13 & 0031905013 & XRT & 2011-06-30 & 3146 & $1.5\pm0.2$ &  \\
14 & 0031905014 & XRT & 2011-07-27 & 3588 & $ 0.9\pm0.1$ &  \\
15 & 0031905015 & XRT & 2011-08-24 &  951 & $1.6\pm0.3$ &  \\
16\tablenotemark{e} & 0031905016 & XRT & 2011-08-29 & 2706 & $1.1\pm0.2$ & \\
17 & 0031905017 & XRT & 2011-09-04 & 4048 & $1.2\pm0.1$ & \\
20 & 0031905018 & XRT & 2012-09-11 & 4882 & $1.2\pm0.1$ &  \\
22 & 0080498001 & XRT & 2013-08-09 & 2187 & $1.0\pm0.1$ & \\
23 & 0080498002 & XRT & 2013-08-21 & 1915 & $1.2\pm0.1$ &  \\
25 & 0080498003 & XRT & 2014-01-20 & 2020 & $ 0.5\pm0.2$ &  \\
\cutinhead{{\it Chandra}}
1A   & 12995      & ACIS-S & 2010-12-23 &  59291 & $1.2\pm0.1$ & $3.6^{+0.2}_{-0.2}$ \\[3pt]
1B   & 13202      & ACIS-S & 2010-12-25 &  98780 & $1.3\pm0.1$ & $4.4^{+0.5}_{-0.4}$ \\[3pt]
10A\tablenotemark{c}  & 12993      & ACIS-S & 2011-03-15 &  49398 & $1.5\pm0.1$ & $4.5^{+0.7}_{-0.3}$ \\[3pt]
10B  & 13241      & ACIS-S & 2011-03-18 &  78963 & $1.6\pm0.1$ & $5.3^{+1.3}_{-0.7}$ \\[3pt]
11A\tablenotemark{d}  & 12994      & ACIS-S & 2011-03-23 & 150058 & $1.6\pm0.1$ & $5.1^{+0.5}_{-0.4}$ \\[3pt]
11B  & 12996      & ACIS-S & 2011-03-29 &  53044 & $1.5\pm0.1$ & $5.3^{+1.1}_{-0.6}$ \\[3pt]
11C  & 13248      & ACIS-S & 2011-04-03 &  54329 & $1.5\pm0.1$ & $5.0^{+1.0}_{-0.5}$ \\[3pt]
16\tablenotemark{e}   & 14332      & ACIS-S & 2011-08-29 &  52381 & $1.0\pm0.1$ & $2.9^{+0.4}_{-0.4}$ \\[3pt]
18   & 12992      & ACIS-S & 2011-09-05 &  66286 & $1.0\pm0.1$ & $3.2^{+0.5}_{-0.5}$ \\[3pt]
19   & 14342      & ACIS-S & 2011-12-28 &  67103 &  $0.8\pm0.1$  & $2.3^{+0.3}_{-0.2}$ \\[3pt]
26   & 16024 & ACIS-I & 2014-06-07 & 29588 & $0.7\pm0.1$  & $2.0^{+0.2}_{-0.1}$ \\
\cutinhead{{\it XMM-Newton}}

21 & 0723450101 & MOS1 & 2013-08-07 & 50143 & $1.15\pm0.02$ & $3.4^{+0.3}_{-0.2}$ \\
 &  & MOS2 &  & 49041 & & \\
 &  & PN &  & 41666 & & \\
24 & 0723450201 & MOS1 & 2014-01-11 & 43060 & $0.20\pm0.01$ & $0.57^{+0.42}_{-0.03}$ \\
 &  & MOS2 & & 42164 & & \\
 &  & PN & & 24922 & & \\
27 & 0729561201 & MOS1 & 2014-07-06 & 27448 & $0.87\pm0.02$ & $2.5^{+0.2}_{-0.2}$ \\
 &  & MOS2 & & 27546 & & \\
 &  & PN & & 22904 & & \\
\enddata
\tablenotetext{a}{Observed flux in the 0.3--10 keV band.}
\tablenotetext{b}{Unabsorbed luminosity in the 0.3--10 keV band, calculated only for observations for which we have modelled the spectrum. For the {\it Chandra} observations, we adopted the luminosities published for the best-fitting {\it diskbb+pow} model, while for the {\it XMM-Newton} observations we used the {\it diskpbb} model.}
\tablenotetext{c}{The \swift\ exposure covered the first part of the \chandra\ exposure;
the \swift\ exposure lasted from 12:17:32 to 22:05:56
while the \chandra\ exposure began at 12:21:40.}
\tablenotetext{d}{The \swift\ exposure covered the end of the \chandra\ exposure;
the \swift\ exposure\ lasted from 11:20:06 to 21:05:57
while the \chandra\ exposure ended at 22:18:33.}
\tablenotetext{e}{The \swift\ and \chandra\ exposures were not quite coincident;
the \swift\ exposure lasted from 11:35:00 to 15:13:56 
while the \chandra\ exposure began at 18:41:51.}
\end{deluxetable*}

\section{X-ray flux and hardness ratio}
The $0.3$--$10$ keV light curve suggests that the outburst peaked 
in February and March 2011, when it displayed strong week-to-week variability 
of as much as factors of two in flux. The flux declined steadily, with 
smaller week-to-week variability through the end of 2011 \citep{soria12}.
Our new observations show that the decline did not continue significantly past 
the end of 2011. Eight measurements in the intervening years show that 
ULX-1 has maintained a roughly constant maximum flux, though with significant 
variations. 
Thus, although the source sometimes fades to below the ULX threshold, 
as it did in January 2014, the bulk of our measurements since 2011 
have caught the source at $\sim2\times10^{39}$ erg s$^{-1}$.

The hardness ratio shown in Figure 2 is the {\it Swift}/XRT count rate 
in the $2$--$10$ keV band divided by the count rate at $0.3$--$2$ keV.
In general, the hardness ratio declines with flux, so that at higher 
luminosities the source is softer than at lower luminosities.
The {\it Chandra} observations in 2010--2011 made it clear that 
a trend in hardness 
ratio was driven primarily by the fraction of the emission in the thermal 
disk (the greater the fraction the softer the overall emission) 
with some modification by the index of the power law.
Most of the more recent data follow the same trend in the hardness-ratio vs 
count-rate diagram.
The clear exception is the {\it XMM-Newton} measurement in January 2014 
(epoch 24) 
when the source was exceptionally faint and the hardness ratio was 
soft rather than hard.
To interpret this possible state transition, 
we will do a full spectral analysis of this observation, and compare it 
with the other two {\it XMM-Newton} observations. 
The {\it Swift} observation made slightly more than a week later 
(epoch 25) has a sufficiently large uncertainty
that its hardness ratio could be consisted with either the low 
{\it XMM-Newton} observation or the previously noted trend.

It is very rare for transient Galactic stellar-mass BHs to be in outburst 
for such a long period of time, $\approx$ 4 years\footnote{See for example 
the {\it Rossi X-ray Timing Explorer} All Sky Monitor lightcurves 
at http://xte.mit.edu/asmlc/ASM.html.}. 
The exception is GRS1915$+$105
which has been in outburst at luminosities $\approx$$10^{39}$ erg s$^{-1}$ 
for over 20 years \citep{castro-tirado94}.
A few others among the most luminous Galactic BHs with a low-mass donor 
have shown long outbursts: 
$\approx$2 years for 4U 1630$-$47 \citep{tomsick05}; 
$\approx$1 year for GRO J1655$-$40 \citep{sobczak99} 
and XTE J1550$-$564 \citep{sobczak00}.
The outburst profile, with its initial peak, decline and reflarings, 
is also at least qualitatively similar to the outbursts of those 
three Galactic BHs.

\begin{deluxetable*}{lccc}
\tablecolumns{4}
\tabletypesize{\scriptsize}
\tablecaption{Spectral fits to the {\it XMM-Newton}/EPIC spectra
\label{tabext:xrayfits}}
\tablewidth{0pt}
\tablehead{
\colhead{Parameter} &
\colhead{Value in 2013 Aug} &
\colhead{Value in 2014 Jan} &
\colhead{Value in 2014 Jul} }
\startdata
\cutinhead{Power law: {\it phabs*phabs*po}}
$n_{\rm H,int}$ ($10^{20}$ cm$^{-2}$) & $12.7^{+0.9}_{-0.9}$ & $17.0^{+2.4}_{-2.2}$ & $11.6^{+1.3}_{-1.3}$\\[3pt]
$\Gamma$          & $1.85^{+0.03}_{-0.03}$    & $2.42^{+0.09}_{-0.09}$  & $1.92^{+0.04}_{-0.04}$\\[3pt]
$N_{\rm po}$ ($10^{-4}$) & $2.57^{+0.08}_{-0.08}$   & $0.81_{-0.06}^{+0.07}$ & $2.06^{+0.10}_{-0.09}$\\[3pt]
$\chi^{2}$/dof & $1.26~(840.2/663)$ & $1.17~(302.96/258)$ & $1.21~(497.54/410)$\\[3pt]
$f_{0.3-10{\rm keV}}$ ($10^{-13}$ erg cm$^{-2}$ s$^{-1}$)  & $12.0^{+0.2}_{-0.2}$ & $2.1^{+0.1}_{-0.1}$ & $9.0^{+0.2}_{-0.2}$\\[3pt]
$L_{0.3-10{\rm keV}}$ ($10^{39}$ erg s$^{-1}$) & $4.0^{+0.1}_{-0.1}$ & $1.00^{+0.07}_{-0.07}$ & $3.1^{+0.1}_{-0.1}$\\


\cutinhead{Cutoff power law: {\it phabs*phabs*cutoffpl}}
$n_{\rm H,int}$ ($10^{20}$ cm$^{-2}$) & $0.4^{+1.4}_{-0.4}$ & $1.4^{+4.3}_{-1.4}$ & $<1.0$\\[3pt]
$\Gamma$  & $0.76^{+0.12}_{-0.07}$ & $0.89^{+0.41}_{-0.27}$ & $0.83^{+0.07}_{-0.07}$\\[3pt]
$E_{\rm efold}$ (keV) & $3.0^{+0.4}_{-0.3}$ & $1.9^{+0.7}_{-0.3}$ & $2.9^{+0.3}_{-0.3}$\\
$N_{\rm po}$ ($10^{-4}$) & $2.38^{+0.07}_{-0.06}$   & $0.85_{-0.07}^{+0.08}$ & $1.98^{+0.08}_{-0.08}$\\[3pt]
$\chi^{2}$/dof & $0.90~(597.99/662)$ & $1.02~(262.63/257)$ & $0.91~(372.04/409)$\\[3pt]
$f_{0.3-10}$ ($10^{-13}$ erg cm$^{-2}$ s$^{-1}$) & $11.5^{+0.2}_{-0.2}$ & $2.0^{+0.1}_{-0.1}$ & $8.7^{+0.2}_{-0.2}$\\[3pt]
$L_{0.3-10{\rm keV}}$ ($10^{39}$ erg s$^{-1}$) & $3.1^{+0.1}_{-0.1}$ & $0.57^{+0.07}_{-0.03}$ & $2.4^{+0.1}_{-0.1}$\\

\cutinhead{Disk blackbody + power law: {\it phabs*phabs*(diskbb+po)}}
$n_{\rm H,int}$ ($10^{20}$ cm$^{-2}$) & $5.7^{+3.7}_{-3.1}$ & $0.2^{+16.0}_{-0.2}$ & $2.4^{+5.0}_{-2.4}$\\[3pt]
$kT_{\rm in}$ (keV)      &  $1.58^{+0.13}_{-0.14}$     & $0.87^{+0.26}_{-0.08}$ & $1.42^{+0.16}_{-0.15}$ \\[3pt]
$N_{\rm dbb}$ ($10^{-3}$) & $6.3^{+2.4}_{-1.4}$   & $13.1_{-9.7}^{+5.9}$ & $7.4^{+3.7}_{-2.4}$\\[3pt]
$\Gamma$  & $2.00^{+0.38}_{-0.28}$  & $1.8^{+1.2}_{-0.6}$ & $1.93^{+0.50}_{-0.33}$\\[3pt]
$N_{\rm po}$ ($10^{-5}$) & $9.6^{+2.4}_{-2.7}$   & $1.1_{-0.4}^{+4.2}$ & $6.0^{+3.1}_{-2.1}$\\[3pt]
$\chi^{2}$/dof & $0.91~(598.15/661)$ & $1.05~(268.21/256)$ & $0.91~(372.17/408)$\\[3pt]
$f_{0.3-10}$ ($10^{-13}$ erg cm$^{-2}$ s$^{-1}$) & $11.5^{+0.2}_{-0.2}$  & $2.0^{+0.1}_{-0.1}$ & $8.7^{+0.2}_{-0.2}$\\[3pt]
$L_{0.3-10{\rm keV}}$ ($10^{39}$ erg s$^{-1}$) & $3.4^{+0.3}_{-0.2}$ & $0.57^{+0.42}_{-0.03}$ & $2.5^{+0.2}_{-0.2}$\\

\cutinhead{Extended disk-blackbody: {\it phabs*phabs*diskpbb}}
$n_{\rm H,int}$ ($10^{20}$ cm$^{-2}$) & $2.9^{+1.4}_{-1.4}$ & $5.0^{+4.5}_{-4.5}$ & $0.8^{+2.1}_{-0.8}$\\[3pt]
$kT_{\rm in}$  (keV) &  $1.90^{+0.12}_{-0.11}$     & $1.23^{+0.22}_{-0.16}$ & $1.72^{+0.17}_{-0.12}$\\[3pt]
$p$ & $0.63^{+0.02}_{-0.02}$ &  $0.56^{+0.07}_{-0.05}$ & $0.63^{+0.02}_{-0.03}$\\[3pt]
$N_{\rm disk}$ ($10^{-3}$) & $2.6^{+1.0}_{-0.8}$   & $1.9_{-1.2}^{+2.7}$ & $3.0^{+1.5}_{-1.2}$\\[3pt]
$\chi^{2}$/dof & $0.90~(595.95/662)$ & $1.03~(263.84/257)$ & $0.91~(370.43/409)$\\[3pt]
$f_{0.3-10{\rm keV}}$ ($10^{-13}$ erg cm$^{-2}$ s$^{-1}$) & $11.5^{+0.2}_{-0.2}$  & $2.0^{+0.1}_{-0.1}$ & $8.6^{+0.2}_{-0.2}$\\[3pt]
$L_{0.3-10{\rm keV}}$ ($10^{39}$ erg s$^{-1}$) & $3.2^{+0.1}_{-0.1}$ & $0.63^{+0.02}_{-0.02}$ & $2.4^{+0.1}_{-0.1}$\\

\cutinhead{Irradiated Comptonized disk: {\it phabs*phabs*diskir}}
$n_{\rm H,int}$ ($10^{20}$ cm$^{-2}$) & $0.5^{+0.8}_{-1.2}$ & $0.1^{+0.8}_{-0.1}$ & $<0.8$\\[3pt]
$kT_{\rm in}$  (keV)     & $0.56^{+0.19}_{-0.16}$   & $0.59^{+0.06}_{-0.13}$ & $0.39^{+0.20}_{-0.15}$ \\[3pt]
$\Gamma$    &  $1.46^{+0.06}_{-0.06}$  & $1.29^{+0.21}_{-\ast}$ & $1.48^{+0.06}_{-0.06}$\\[3pt]
$kT_e$  (keV)      & $1.44^{+0.22}_{-0.17}$   & $>8.3$ & $1.29^{+0.16}_{-0.16}$\\[3pt]
$L_c/L_d$    &  $1.9_{-0.5}^{+2.8}$  & $5.0^{+3.5}_{-2.0}$ & $2.7^{+3.3}_{-1.5}$ \\[3pt]
$N_{\rm dbb}$ ($10^{-3}$) & $200^{+45}_{-15}$   & $53_{-23}^{+62}$ & $473^{+1450}_{-360}$\\[3pt]
$\chi^{2}$/dof    & $0.91~(597.84/660)$   & $1.03~(263.43/255)$ & $0.91~(371.34/407)$\\[3pt]
$f_{0.3-10{\rm keV}}$ ($10^{-13}$ erg cm$^{-2}$ s$^{-1}$)  & $11.5^{+0.2}_{-0.2}$   & $2.0^{+0.1}_{-0.1}$ & $8.6^{+0.2}_{-0.2}$\\[3pt]
$L_{0.3-10{\rm keV}}$ ($10^{39}$ erg s$^{-1}$) & $3.1^{+0.1}_{-0.1}$ & $0.56^{+0.03}_{-0.02}$ & $2.3^{+0.1}_{-0.1}$\\
\enddata
\tablenotetext{a}{All errors are given at the 90\% confidence level. In all cases, we assume a fixed Galactic column density $n_{\rm H} = 4 \times 10^{20}$ cm$^{-2}$ in addition to a fitted intrinsic absorption column. Fluxes are observed values; luminosities are defined as $4\pi d^2 \times$ the unabsorbed flux. For the {\it diskir} model, we assumed $f_{in} = 0.1$, $r_{irr} = 1.2$, $f_{out} = 0.005$ and $\log r_{\rm out} = 5.0$.}
\end{deluxetable*}

\section{X-ray spectral state}
To better understand the outburst evolution, 
we carried out spectral fits to the {\it XMM-Newton} data. 
The first question we asked was whether the spectra are consistent 
with a straight power-law, or are significantly curved. 
There is still no consensus on a definition of ``canonical'' ULX 
spectral states, but it has been clear from the earliest studies 
that, as a zeroth order approximation, some ULXs are in a power-law-like 
state, others have a curved spectrum, and some switch between the two states 
\citep{kubota01,makishima07,soria11}.
As shown in Table 2, we find that none of the three {\it XMM-Newton} spectra 
is well fitted with a simple absorbed power-law; 
all exhibit significant curvature.
An exponentially cut-off power-law ({\it cutoffpl} model in {\small {XSPEC}})
is a phenomenological way of highlighting the high-energy curvature, 
and this model does indeed provide very good fits (Table 2). The slope 
(photon index) of the power-law below the exponential cut-off is 
$\Gamma \approx 0.8$, much too flat for any physically plausible 
model ({\it e.g.}, inverse-Compton or synchrotron emission) 
applicable to X-ray binaries in the low/hard state. 
It corresponds to a flux density $F_\nu \propto \nu^{0.2 \pm 0.1}$, which is 
more consistent with the ``flat'' part of an accretion disk just below 
the Wien cutoff \citep{ss73,frank02}.

We then tried more physical models, suitable to luminous X-ray binaries. 
A disk-blackbody \citep{ss73,mitsuda84,makishima86} 
plus power-law is the standard model used in the literature 
for classifying the accretion state of X-ray binaries. 
This model provides good fits to all three epochs 
of {\it XMM-Newton} data (Table 2). However, while such models formally 
describe 
the data, it is important to examine whether the values of the best-fitting 
parameters are physically plausible or self-consistent.
In the first and third epoch, we find 
inner-disk temperatures $\approx 1.5$ keV, which are too high 
for a standard accretion disk around stellar-mass BHs. 
In Galactic BHs, typical values of the inner-disk 
temperature in the disk-dominated high/soft state are $\la 1.2$ keV 
\citep{mcclintock06}. Given its high luminosity, M\,83 ULX-1 
might be (if anything) a little more massive than 
typical Galactic BHs, in which case its disk temperature should be even lower.
Only the second epoch of our {\it XMM-Newton} spectra is consistent 
with a canonical high/soft state dominated by a standard disk-blackbody 
with $kT_{\rm in} \approx 0.9$ keV.

Higher-than-expected disk color temperatures are observed 
in other ULXs \citep{stobbart06,roberts07}, and have been interpreted 
in two ways (although the difference may be at least partly semantic).
In one model \citep{watarai01,mizuno01,kubota04,isobe12}, 
the standard disk evolves into a slim disk 
when the accretion rate reaches a critical level and the luminosity approaches 
or mildly exceeds the Eddington limit. 
One of the observable properties that distinguish 
a slim disk from a standard disk is a flatter radial temperature profile 
($T(R) \propto R^{-p}$, where $p \approx 0.5--0.6$ rather than $p=0.75$ as 
in the standard disk case). Another one is non-negligible emission 
from inside the innermost 
stable circular orbit \citep{watarai00,mizuno01,kulkarni11}, 
which makes the inner disk radius appear smaller 
than in the standard case. The slim disk model predicts higher peak color 
temperature ($kT_{\rm in} \approx 1.5$--$2$ keV), as the inner 
part of the disk becomes dominated by electron scattering and radiative 
emission becomes less efficient.
Alternatively \citep{done06,gladstone09,sutton14}, 
the same near-Eddington regime can be modelled 
as a warm ($kT_{e} \approx 1.5$--$2$ keV), optically thick ($\tau \sim 10$) 
scattering corona covering or replacing the inner part of a standard disk. 
The observed spectrum is then a combination of a disk-blackbody from 
the uncovered (larger and cooler) outer disk plus a scattering component 
from the warm corona, with a downturn above 3 keV.
Both scenarios have been applied 
not only to ULXs, but also to the highest luminosity 
phases of some Galactic BHs; for example, the peak of the outburst in 
XTE J1550$-$564 was successfully explained in the slim-disk framework 
\citep{kubota04} as well as in the warm-corona scenario \citep{done06}.
In some cases, a second Comptonization component, produced in a hotter 
($kT_{e} > 10$ keV), thinner corona ($\tau \sim 1$) 
may also be present, and responsible for the power-law component.

In our {\small {XSPEC}} spectral analysis of M\,83 ULX-1, 
we used the {\it diskpbb} model 
as an approximation of the slim disk (Figures 3, 4). This is also known 
in the literature as a p-free disk, or extended multicolor disk \citep{isobe12}.
We used the Comptonization model {\it diskir} \citep{gierlinski09} 
to reproduce the outer-disk plus warm-corona scenario.
Both sets of models provide formally good fits (Table 2). 
Some of the parameters 
in the Comptonization model for the 2014 January spectrum are poorly 
constrained because there is little evidence of a straight power-law 
above the curved thermal component at high energies.
In fact, adding a power-law component to the {\it diskpbb} model 
does not improve the fit.

There is little difference in the fit statistics 
between the interpretation of our new results as a slim-disk regime 
(or other similar types of modified disk), or as a standard-outer-disk 
plus warm, optically-thick Comptonized component; thus, we must look 
at the general evolution of the source from epoch to epoch 
to make a tentative choice between them.
We had interpreted the 2010--2011 {\it Chandra}/ACIS spectra 
in the frame of a disk plus hot-corona model \citep{soria12}
because there is no evidence of a high-energy downturn or curvature 
above 2 keV in the {\it Chandra} spectra, which were mostly taken 
at higher luminosity phases of the outburst. 
For all those spectra, we can only place a lower limit 
to the coronal temperature, as $kT_e > 1.8$ keV at all epochs  
\citep{soria12} and in one case (2011 March 15) $kT_e > 6.5$ keV. 
However, estimating or constraining a possible 
high-energy downturn in those spectra is difficult for 
at least two reasons: firstly, {\it Chandra}'s sensitivity above 5 keV is much 
lower than {\it XMM-Newton}'s; secondly, the {\it Chandra} spectra suffer 
from pile-up, which flattens the slope at high energies and may 
mask a downturn. Thus, we suggest now that ULX-1 was more dominated 
by Comptonization during the {\it Chandra} observations, 
but we cannot make any stronger inference on 
the temperature of the Comptonizing medium (see Section 6 
for a more detailed discussion of the differences between the spectral 
states in the two sets of observations).

In the {\it XMM-Newton} spectra, instead, 
the high-energy curvature is significant, 
and corresponds to characteristic temperatures  
$kT \approx$ 1.5 keV, 1.2 keV and 1.7 keV for the three epochs.
Taking into account that these temperatures are relatively low 
for a Comptonizing corona, and that the second epoch is also consistent 
with a canonical high/soft state, we favour the slim-disk 
interpretation for these observations. We suggest that in 2013--2014, ULX-1 
was at the boundary between the high/soft and slim disk states, 
varying between just below and just above the Eddington luminosity.
The trend of increasing temperatures with flux suggests 
disk emission as the most plausible interpretation. For coronal emission, 
we would instead expect a decrease in temperatures with increasing flux, 
because the incresed illumination would cool the corona \citep{middleton11}. 
The low value of $p \approx 0.6 < 0.75$ found in all three epochs indicates 
that the disk is always non-standard, including during the 2014 January 
observation, when the X-ray luminosity was $\approx 6 \times 10^{38}$ 
erg s$^{-1}$ (Table 2). Extensive studies of bright Galactic BH X-ray 
binaries show that deviations from the standard disk spectrum start 
to appear at disk luminosities $\sim 0.3 L_{\rm Edd}$ \citep{steiner10}.
This suggests that the BH mass of M\,83 ULX-1 is $\la 20 M_{\odot}$.
For a slim disk model, we would have expected $p$ to decrease with 
luminosity, with $p \rightarrow 0.5 $ at the highest fluxes; instead, 
the index is consistent with a constant value $p \approx 0.6$ 
at all three epochs. It is unclear how to interpret this finding.  
It is important to remember that X-ray spectral fits give us the scaling of 
the color temperature $T_{\rm col} \propto R^{-p}$, with 
$T_{\rm col}$ related to the effective temperature $T_{\rm eff}$ by a hardening 
factor $\kappa$, that is $T_{\rm col} \equiv \kappa T_{\rm eff}$ 
\citep{shimura95}. A small dependence 
of $\kappa$ on disk radius at the highest accretion rates  
would be sufficient to produce the observed small discrepancy from 
the underlying slim-disk relation $T_{\rm eff} \propto R^{-0.5}$.
Therefore, we do not think that the fitted values of $p$ are  
a significant argument against a disk-dominated spectrum.

In terms of unaborbed luminosities, regardless of the spectral model, 
ULX-1 was below the formal ULX threshold in 2014 January, 
while it was in the ULX regime in the other two epochs, although 
a factor of 2 or 3 fainter than in early 2011. 
Note that, for simplicity, all luminosities listed in Table 2 
are defined as $4\pi d^2 \times$ the unabsorbed flux (isotropic emission). 
Strictly speaking, this is a good approximation for the emission 
from a spherical or quasi-spherical inflow ({\it e.g.}, 
the power-law component), while for a standard thin disk, 
the luminosity is $2\pi d^2 (\cos \theta)^{-1} \times$ 
the flux, where $\theta$ is the (unknown) viewing angle. 
For the slim disk model, we also numerically calculated 
the isotropic bolometric luminosity in the three epochs: 
$L_{\rm bol} = (3.7 \pm 0.1) \times 10^{39}$ erg s$^{-1}$ in 2013 August, 
$(9.2 \pm 0.2) \times 10^{38}$ erg s$^{-1}$ in 2014 January, and
$(2.7 \pm 0.1) \times 10^{39}$ erg s$^{-1}$ in 2014 July.
If our interpretation of ULX-1 as a source 
that is straddling the Eddington threshold is correct 
(see Section 5), these 
luminosities correspond to BH masses $\sim$$10$--$20 M_{\odot}$, 
consistent with a normal stellar-mass BH.

\begin{figure}[t]
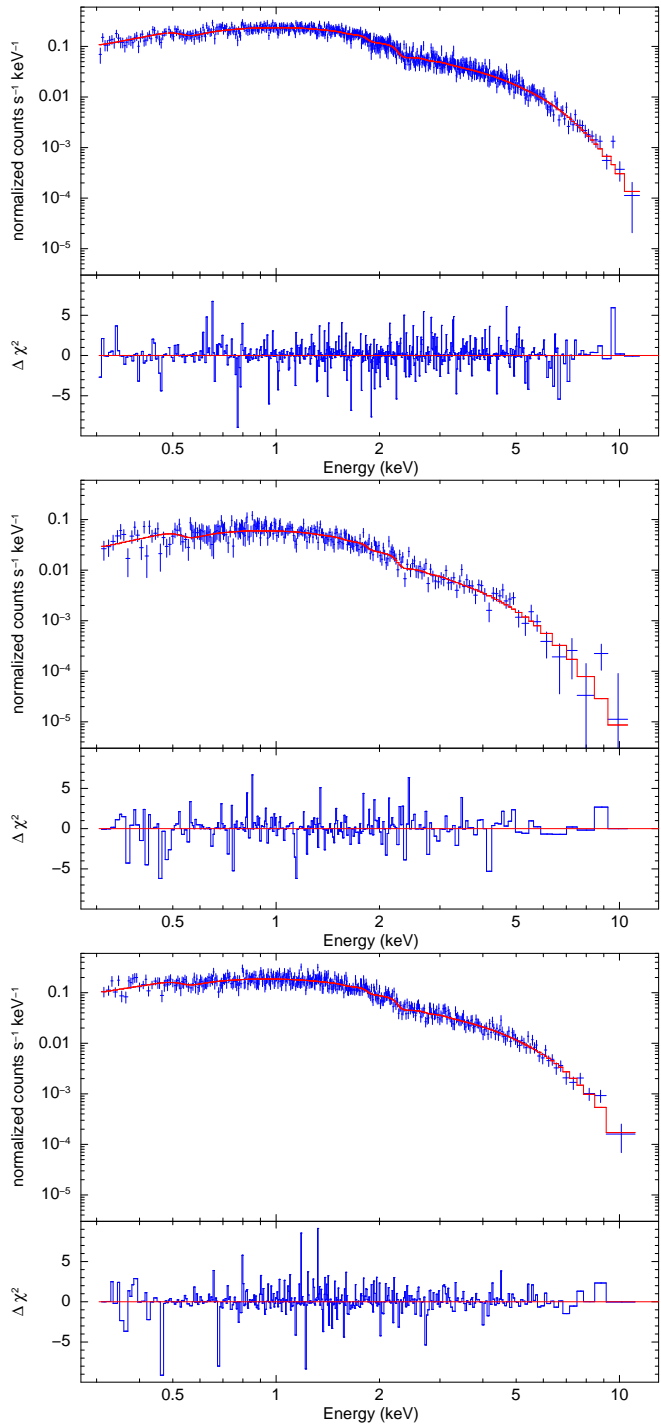

\includegraphics[angle=270,scale=.36]{f3a.ps}\\
\includegraphics[angle=270,scale=.36]{f3b.ps}\\
\includegraphics[angle=270,scale=.36]{f3c.ps}\\
\caption{Top panel: {\it XMM-Newton}/EPIC spectrum and $\chi^2$ residuals 
from the 2013 August observation, fitted with an extended (p-free) 
disk-blackbody model ({\it diskpbb} in {\small {XSPEC}}). 
We show here the average EPIC spectrum obtained 
by combining pn and MOS spectral files with the {\small {SAS}} task 
{\it{epicspeccombine}}. 
Middle panel: same, for the 2014 January observation. 
Bottom panel: same, for the 2014 July data. See Table 2 
for the best-fitting parameters.
\label{fig3}}
\end{figure}

\section{BH mass estimate}
When disk emission dominates the spectrum of an accreting BH, 
it is possible to determine the physical inner disk radius $R_{\rm in}$ 
from the {\small{XSPEC}} model normalization $N$, 
provided that we can constrain two correction factors: the hardening factor 
$\kappa$ for the observed color temperature, and a geometric factor $\xi$ 
that depends on how close to the innermost stable circular orbit 
the disk reaches its peak temperature:
\begin{equation} 
R_{\rm in} \approx \xi \kappa^2 N^{1/2} (\cos \theta)^{-1/2} d_{\rm 10kpc} \ \ {\mathrm {km}} 
\end{equation}
\citep{kubota98,makishima00}.
Then, from $R_{\rm in}$ we can estimate a characteristic BH mass, as a function 
of spin parameter: $R_{\rm in} \approx \alpha GM_{\rm BH}/c^2$, 
where $\alpha = 6$ for a non-rotating BH, $\alpha = 1$ for a maximally 
rotating Kerr BH, and $\alpha = 1.24$
for the maximum spin achievable by an astrophysical BH (spin parameter 
$a = 0.998$; \citealt{thorne74}).

In the case of standard disks, $\xi \approx 0.412$ \citep{kubota98} 
and $\kappa \approx 1.7$ \citep{shimura95,davis05}, so that 
$R_{\rm in} \approx 1.19 \times 460 N^{1/2} (\cos \theta)^{-1/2}$ km 
at the adopted distance of M\,83. 
If we interpret the 2014 January spectrum as a canonical high/soft state, 
we obtain $R_{\rm in} \approx 63^{+15}_{-32} (\cos \theta)^{-1/2}$ km (Table 2).
This is consistent with a slowly-spinning 10-$M_{\odot}$ BH, 
or a maximally-spinning 60-$M_{\odot}$ BH.

However, this approximation holds only for luminosities 
$\la 0.3 L_{\rm Edd}$. At higher accretion rates, the hardening factor 
$\kappa$ increases with luminosity and can be as high as $\approx 2.5$--$3$
for some Galactic BHs ({\it e.g.}, GRO J1655$-$40 at outburst peak, 
and GRS 1915$+$105) and for ULXs in the slim-disk state 
\citep{watarai03,kawaguchi03,shrader03,isobe12}. 
This is why at near-Eddington luminosities, the fitted color temperature 
$T_{\rm in}$ increases faster than the standard 
$T_{\rm in} \propto L_{\rm disk}^{1/4}$ high/soft state relation 
(``anomalous regime'': \citealt{kubota04,abe05}). 
Given the luminosities of M\,83 ULX-1 in the three {\it XMM-Newton} 
epochs, and the fact that a disk model with $p \approx 0.6$ 
provides the best fit to the data, we argue that the source was likely to be 
in the anomalous regime (slightly below Eddington) or in the slim disk regime 
(slightly above Eddington), and therefore we take a hardening factor 
$\kappa \approx 3$ for BH mass estimates. Following \citet{vierdayanti08}, 
we also take $\xi \approx 0.353$, which takes into account the transonic flow 
in the pseudo-Newtonian potential.
Taking an average normalization constant 
$N_{\rm disk} = (2.5 \pm 0.5)\times 10^{-3}$ from our {\it diskpbb} fit 
(Table 2), we obtain a ``true'' inner disk radius 
$R_{\rm in} \approx (73 \pm 15) (\cos \theta)^{-1/2}$ km.
This corresponds to an ``apparent'' BH mass  
\begin{equation}
M_{\rm X} = \left(\frac{6}{\alpha}\right)
\left(\frac{\xi}{0.353}\right) \left(\frac{\kappa}{3}\right)^2  
\frac{8.2 \pm 1.7}{(\cos \theta)^{1/2}} \, M_{\odot}.
\end{equation}
Finally, we need to take into account that the inner radius of a slim disk 
extends slightly inside the innermost stable circular orbit, 
so that the true mass $M_{\rm BH} \approx 1.2 M_{\rm X}$ 
\citep{vierdayanti08}. This gives our final best estimate of the BH mass 
as 
\begin{equation}
M_{\rm BH} = \left(\frac{6}{\alpha}\right)
\left(\frac{\xi}{0.353}\right) \left(\frac{\kappa}{3}\right)^2  
\frac{10 \pm 2}{(\cos \theta)^{1/2}} \, M_{\odot}. 
\end{equation}
In principle, the estimated mass can be as high as $\approx$$60 M_{\odot}$ 
in the (implausible) extreme Kerr scenario; however, the fact that the system 
appears to be in the anomalous regime (upper end of the high/soft state) 
or in the slim disk state for a moderate bolometric luminosity 
$\approx 2 \times 10^{39}$ erg s$^{-1}$ (and possibly even lower, 
considering the 2014 January spectrum) suggests that 
the BH mass is closer to $\sim$$10$--$20 M_{\odot}$.
The same argument holds if we intepret the X-ray spectra as Comptonized 
emission from a warm, optically-thick corona, which is typical of sources 
at $L_{\rm X} \approx 1$--$3 L_{\rm Edd}$ \citep{gladstone09}. 

Our newly estimated BH mass $\sim$$10$--$20 M_{\odot}$ is less than 
the mass estimate proposed in \citet{soria12}. 
In that study, we had not observed the source in a thermal state, 
which limited our ability to estimate the inner disk radius; 
moreover, in the absence of other constraints, 
we conservatively assumed that ULX-1 peaked at around
its the Eddington limit. Instead, if our revised mass estimate is correct, 
ULX-1 must have peaked at $L_{\rm X} \approx 3$--$4L_{\rm Edd}$ during the 2010--2011 
{\it Chandra} observations.

\begin{figure}[t]
\includegraphics[angle=270,scale=.36]{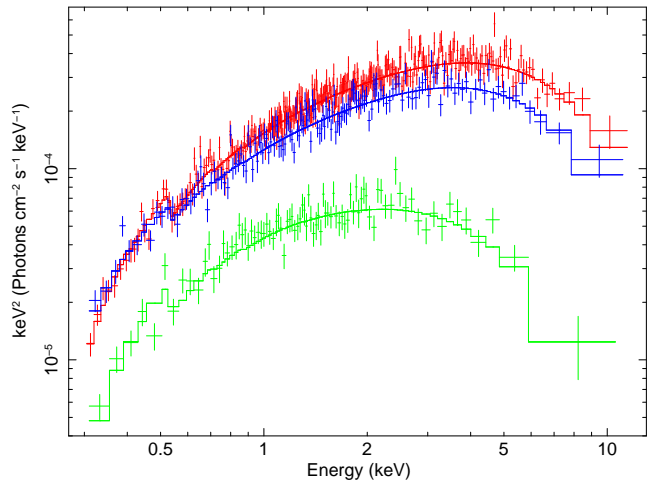}\\
\caption{Unfolded {\it XMM-Newton}/EPIC spectra 
for all three epochs, fitted with the {\it diskpbb} model 
in {\small {XSPEC}}; 2013 August = red; 2014 January = green; 
2014 July = blue. The spectra have been rebinned to a minimum signal-to-noise 
ratio of 6 in each bin, for display purposes only. See Table 2 
for the best-fitting parameters.
\vspace{0.3cm}
\label{fig4}}
\end{figure}

\section{Super-Eddington regime}
Although the idea of moderately super-Eddington emission was somewhat 
frowned upon until recently, there is no strong theoretical argument 
against it: it is well known that above the critical accretion rate, 
$L \approx L_{\rm Edd} (1 + a \ln \dot{m})$ where $0.6 \la a \la 1$ 
\citep{ss73,poutanen07,king14}.
Hence, the accretion rate required 
to produce $L_{\rm X} \approx 3L_{\rm Edd}$ is $\dot{m} \sim 10$ 
(regardless of BH mass).
For a 15-$M_{\odot}$ BH, with $L_{\rm Edd} \approx 2 \times 10^{39}$ erg s$^{-1}$, 
$\dot{m} \sim 10$ corresponds to a physical accretion rate 
$\dot{M} \sim 3 \times 10^{-6} M_{\odot}$ yr$^{-1}$, 
which is a plausible value if the Roche-lobe-filling 
donor star is expanding along the red giant branch or undergoing 
asymptotic-giant-branch pulsations. 
Observationally, several recent studies  
have shown evidence of super-critical accretion and in some cases, 
mildly super-Eddington luminosity \citep{liu13,motch14,soria14}.

A more complicated and unsolved problem is determining when 
super-Eddington BHs have a thermal, curved X-ray spectrum 
(slim-disk or warm, optically-thick corona models) and when they have 
instead a power-law-dominated spectrum 
(consistent with an hotter, optically-thin corona).
Some ULXs ({\it e.g.}, IC\,342 X-1 and X-2: \citealt{kubota01}; 
NGC\,1313 X-2: Gris\'{e} et al., in prep.) 
have been observed in a power-law-like state at lower luminosities, 
and in a curved-spectrum state at higher luminosities.
However, other ULXs ({\it e.g.}, Holmberg IX X-1: \citealt{soria11} 
and references therein) have been observed to switch between power-law 
and curved spectra over a largely overlapping range of luminosities.
M\,83 ULX-1 was certainly brighter during the {\it Chandra} 
observations, when its spectrum was more power-law dominated 
without evidence of a high-energy downturn but with a soft excess at 
low temperatures ($kT \sim 0.2$--$0.4$ keV).

We have argued here that the value of $R_{\rm in} \sim 70$ km inferred 
from the {\it XMM-Newton} spectra is roughly representative 
of the true inner-disk radius (allowing for small correction factors). 
We have also shown that the fitted temperature increased with luminosity.
The {\it Chandra} spectra were power-law-dominated but they also 
included a thermal component, and they were successfully fitted with 
a power-law plus disk-blackbody model; however, in that case, 
the characteristic radius $R_{\rm c}$ inferred from the 
disk-blackbody component was $R_{\rm c} \sim 700$--$1000$ km \citep{soria12}.
and the fitted temperature was lower, despite the higher luminosity.
Our preferred explanation for this difference is that we are not measuring 
the same physical structure. In the {\it XMM-Newton} observations, taken  
near or just above the Eddington limit, the disk-like curvature 
of the spectrum and its high peak temperature ($kT \sim 1$--$2$ keV) 
suggest that we are seeing emission directly from the inner region 
of the disk, close to the innermost stable circular orbit. 
In the {\it Chandra} observations, which we now recognize (by comparison 
with the {\it XMM-Newton} data) to correspond to higher accretion rates, 
the emission from the inner disk is completely Comptonized into a power-law, 
and the soft thermal excess corresponds to emission from larger disk radii, 
outside the hot Comptonizing region. This is consistent with its lower 
temperature ($kT \sim 0.2$--$0.4$ keV) and lower relative contribution 
with respect to the power-law component. 
For power-law-dominated ULXs with a soft excess, it was suggested 
that the fitted radius of the thermal component may correspond 
to the spherization radius, so that $R_c \sim \dot{m} R_{\rm in}$ 
\citep{soria07,kajava09}. This suggests that the {\it XMM-Newton} observations 
($\dot{m} \sim 1$) reveal the true inner-disk radius, 
while the {\it Chandra} observations ($\dot{m} \sim 10$) give 
a characteristic radius an order of magnitude larger.
Mis-identification of this larger characteristic radius 
in power-law-dominated ULXs as the innermost stable circular orbit 
led to speculations (now generally thought to be incorrect) that 
many ULXs could be intermediate-mass BHs with 
masses of a few $10^3 M_{\odot}$ \citep{miller04,soria07}.
 
It is instructive to compare M\,83 ULX-1 with M\,33 X-8 
(the brightest ULX in the Local Group), whose spectral properties 
and variability have been extensively studied with {\it XMM-Newton} 
\citep{middleton11} and {\it Suzaku} \citep{isobe12}.
The spectrum of M\,33 X-8 displays a similar degree of curvature, 
and has been modelled either as slim disk
or as a two-component inflow (warm, optically thick Comptonization medium 
in the inner region plus a cooler outer disk).
The Comptonization model was preferred by \cite{middleton11} because 
the characteristic temperature of M\,33 X-8 decreases with flux (contrary 
to the expectations for a slim disk) and a high-energy tails appears 
at the highest fluxes. In the {\it XMM-Newton} observations 
of M\,83 ULX-1, instead, the characteristic 
temperature increases with flux (Table 2) and there is no hint 
of high-energy tails.
Therefore, a slim-disk model is self consistent. For the {\it Chandra}
spectra of M\,83 ULX-1, the two-component Comptonization model provides 
the most physical explanation.
In summary, as a rule of thumb, we interpret the curved spectrum 
of a ULX as disk emission when its fitted temperature increases with flux, 
and as Comptonization plus outer disk emission when its fitted temperature 
decreases with flux.


\cite{sutton13} have proposed a more refined three-state ULX classification, 
which distinguishes between 
``broadened disk'', ``hard'', and ``soft'' ultraluminous spectra, 
roughly in order of increasing accretion rate and increasing optical depth 
of the outflows. Both the broadened-disk ($1 \la \dot{m} \la 10$) 
and the soft state (corresponding to the highest range of $\dot{m}$) 
appear curved, with a high-energy downturn, 
while the hard state is more power-law-like.
In this picture, M\,83 ULX-1 may have been 
in the broadened-disk regime (just above Eddington) 
during the {\it XMM-Newton} observations, 
and in the hard ultraluminious regime during the {\it Chandra} observations.
In addition to the accretion rate, 
it is likely that other factors such as the viewing angle to the disk,  
wind geometry and hysteresis determine whether a ULX appears to us 
as a power-law or curved X-ray spectrum.
\cite{kawashima12} have carried out hydrodynamic simulations 
which show how Comptonization and downscattering in a thick outflow 
can provide a physical explanation 
for the evolution between power-law and curved ULX spectra 
as a function of accretion rate.

\section{Conclusions}
We used {\it XMM-Newton}, {\it Swift} and {\it Chandra} 
to follow the evolution of the transient ULX in M\,83 that 
went into outburst sometime between late 2009 and 2010. 
Four years later, the outburst continues, and M\,83 ULX-1  
remains the most luminous point-like object in that galaxy, 
although not as bright as in 2011 February--March.
After a temporary decline , during which X-1 dipped below 
the ULX threshold, it has now returned to an X-ray luminosity 
$\approx 2 \times 10^{39}$ erg s$^{-1}$. 

Using high-quality {\it XMM-Newton} spectra from 2013--2014, 
we showed that M\,83 ULX-1 has a curved spectrum, consistent with 
a slim disk regime, or more generally, with disk emission 
heavily modified by optically-thick Comptonization in a warm medium 
($kT_e \la 2$ keV).
From the best-fitting model parameters in the three {\it XMM-Newton} 
epochs, we argue that the source is 
currently varying between the upper end of the high/soft state 
and a mildly super-Eddington (ultraluminous) state. 
Thus, the evolution of this source helps us understand the close relation 
between the most luminous Galactic BH binaries and ULXs.

From the best-fitting radius and temperature of the inner disk 
(in the framework of the slim-disk model), 
and from our argument that the source luminosity is varying within a factor 
of 2 of an Eddington luminosity $\approx 2 \times 10^{39}$ erg s$^{-1}$,
we infer a BH mass $\approx10$--$20 M_{\odot}$ 
(subject to uncertainties in the viewing angle and BH spin). 
This is consistent with the maximum mass of stellar BHs we expect to find 
in a metal-rich galaxy such as M\,83. It is a further argument in favour 
of the interpretation of most ULXs as super-critical accretors.

We plan to continue monitoring the outburst evolution of M\,83 ULX-1.
If the outburst is declining, we will try to determine at what luminosities 
the system switches to the canonical high/soft state and then 
to the low/hard state, and therefore get a more accurate estimate 
of its BH mass and of its Eddington luminosity. 
If the outburst re-flares, we will test whether ULX-1 switches again 
to a power-law-dominated spectrum (as seen in the 2010--2011 {\it Chandra} 
observations) at higher luminosity.

\acknowledgments
We thank our additional colleagues on the companion multi-wavelength surveys of M\,83 for useful discussions. We thank the anonymous referee for useful suggestions that have improved our discussion. R.S. also thanks Chris Done, Hua Feng, Fabien Gris\'{e}, Matthew Middleton, Tim Roberts for useful discussions on ULX spectral states relevant to this work.
Support for this work was provided by the National Aeronautics and Space Administration through Chandra grant No. GO1-12115, issued by the Chandra X-Ray Observatory Center, which is operated by the Smithsonian Astrophysical Observatory for and on behalf of NASA under contract NAS8-03060. 
R.S. acknowledges an Australian Research Council grant DP120102393.
W.P.B. and K.K. acknowledge Chandra grant No.~GO1-12115C to Johns Hopkins University. P.F.W. also acknowledges financial support from the National Science Foundation through grant AST-0908566.


\begin{thebibliography}{}
\bibitem[Abe et al.(2005)]{abe05} Abe, Y., Fukazawa, Y., Kubota, A., Kasama, D., \& Makishima, K. 2005, PASJ, 57, 629
\bibitem[Arnaud(1996)]{arnaud96} Arnaud, K. A. 1996, Astronomical Data Analysis Software and Systems V (ASP Conf. Ser. 101), ed. G. H. Jacoby \& J. Barnes (San Francisco, CA: ASP), 17
\bibitem[Belczynski et al.(2010)]{belczynski10} Belczynski, K., Bulik, T., Fryer, C. L., Ruiter, A., Valsecchi, F., Vink, J. S., \& Hurley, J. R. 2010, ApJ, 714, 1217
\bibitem[Blackburn(1995)]{blackburn95} Blackburn, J. K. 1995, Astronomical Data Analysis Software and Systems IV (ASP Conf. Ser. 77), ed. R. A. Shaw, H. E. Payne, \& J. J. E. Hayes (San Francisco, CA: ASP), 367
\bibitem[Bresolin \& Kennicutt(2002)]{bresolin02} Bresolin, F., \& Kennicutt, R. C. 2002, ApJ, 572, 838
\bibitem[Castro-Tirado et al.(1994)]{castro-tirado94} Castro-Tirado, A. J., Brandt, S., Lund, N., Lapshov, I., Sunyaev, R. A., Shlyapnikov, A. A., Guziy, S., \& Pavlenko, E. P. 1994, ApJS, 92, 469
\bibitem[Davis et al.(2005)]{davis05} Davis, S. W., Blaes, O. M., Hubeny, I., \& Turner, N. J. 2005, ApJ, 621, 372
\bibitem[Done \& Kubota(2006)]{done06} Done, C., \& Kubota A. 2006, MNRAS, 371, 1216
\bibitem[Fender et al.(2004)]{fender04} Fender, R. P., Belloni, T. M., \& Gallo, E. 2004, MNRAS, 355, 1105
\bibitem[Feng \& Soria(2011)]{feng11} Feng, H., \& Soria, R. 2011, NewA Rev., 55, 166
\bibitem[Frank et al.(2002)]{frank02} Frank, J., King, A. R., \& Raine, D. J. 2002, Accretion Power in Astrophysics (Cambridge: Cambridge University Press)
\bibitem[Fruscione et al.(2006)]{fruscione06} Fruscione, A., McDowell, J. C., 
Allen, G. E., et al. 2006, Proc. SPIE, 6270, 62701V
\bibitem[Gierli\'{n}ski et al.(2009)]{gierlinski09} Gierli\'{n}ski, M., 
Done, C., \& Page, K. 2009, MNRAS, 392, 1106
\bibitem[Gladstone et al.(2009)]{gladstone09} Gladstone, J. C., Roberts, T. P., \& Done, C. 2009, MNRAS, 397, 1836
\bibitem[Isobe et al.(2012)]{isobe12} Isobe, N., Kubota, A., Sato, H., \& Mizuno, T. 2012, PASJ, 64, 119
\bibitem[Kajava \& Poutanen(2009)]{kajava09} Kajava, J. J. E., \& Poutanen, J. 2009, MNRAS, 398, 1450
\bibitem[Kawaguchi(2003)] {kawaguchi03} Kawaguchi, T. 2003, ApJ, 593, 69
\bibitem[Kawashima et al.(2012)]{kawashima12} Kawashima, T., Ohsuga, K., Mineshige, S., Yoshida, T., Heinzeller, D., \& Matsumoto, R. 2012, ApJ, 752, 18
\bibitem[King(2014)]{king14} King, A. R. 2014, Science, 343, 1318
\bibitem[Kreidberg et al.(2012)]{kreidberg12} Kreidberg, L., Bailyn, C. D., Farr, W. M., \& Kalogera, V. 2012, ApJ, 757, 36
\bibitem[Kubota \& Makishima(2004)]{kubota04} Kubota, A., \& Makishima, K. 2004, ApJ, 601, 428
\bibitem[Kubota et al.(2001)]{kubota01} Kubota, A., Mizuno, T., Makishima, K., Fukazawa, Y., Kotoku, J., Ohnishi, T., \& Tashiro, M. 2001, ApJ, 547, L119
\bibitem[Kubota et al.(1998)]{kubota98} Kubota, A., Tanaka, Y., Makishima, K., Ueda, Y., Dotani, T., Inoue, H., \& Yamaoka, K. 1998, PASJ, 50, 667
\bibitem[Kulkarni et al.(2011)]{kulkarni11} Kulkarni, A. K., et al. 2011, MNRAS, 414, 1183
\bibitem[Liu et al.(2013)]{liu13} Liu, J.-F., Bregman, J. N., Bai, Y., Justham, S., \& Crowther, P. 2013, Nature, 503, 500
\bibitem[Long et al.(2014)] {long14} Long, K. S., Kuntz, K. D., Blair, W. P., Godfrey, L., Plucinsky, P. P., Soria, R., Stockdale, C., \& Winkler, P. F. 2014, ApJS, 212, 21
\bibitem[McClintock \& Remillard(2006)] {mcclintock06} McClintock, J. E., \& Remillard, R. A. 2006, Black Hole Binaries, ed. W. H. G. Lewin \& M. van der Klis (Cambridge: Cambridge Univ. Press), 157
\bibitem[Makishima(2007)]{makishima07} Makishima, K. 2007, IAUS, 238, 209
\bibitem[Makishima et al.(1986)]{makishima86} Makishima, K., Maejima, Y., Mitsuda, K., Bradt, H. V., Remillard, R. A., Tuohy, I. R., Hoshi, R., \& Nakagawa, M. 1986, ApJ, 308, 635
\bibitem[Makishima et al.(2000)]{makishima00} Makishima, K., et al. 2000, ApJ, 535, 632
\bibitem[Middleton et al.(2011)]{middleton11} Middleton, M. J., Sutton, A. D., \& Roberts, T. P. 2011, MNRAS, 417, 464
\bibitem[Miller et al.(2004)]{miller04} Miller, J. M., Fabian, A. C., \& Miller, M. C. 2004, ApJ, 614, L117
\bibitem[Mitsuda et al.(1984)]{mitsuda84} Mitsuda, K., et al. 1984, PASJ, 36, 741
\bibitem[Mizuno et al.(2001)]{mizuno01} Mizuno, T., Kubota, A., \& Makishima, K. 2001, ApJ, 554, 1282	
\bibitem[Motch et al.(2014)]{motch14} Motch, C., Pakull, M. W., Soria, R., Gris\'{e}, F., \& Pietrzy\'{n}ski, G. 2014, Nature, 514, 198
\bibitem[Pilyugin et al.(2006)]{pilyugin06} Pilyugin, L. S., Thuan, T. X., \& V\'{i}lchez, J. M. 2006, MNRAS, 367, 1139
\bibitem[Pilyugin et al.(2010)]{pilyugin10} Pilyugin, L. S., V\'{i}lchez, J. M., \& Thuan, T. X. 2010, ApJ, 720, 1738
\bibitem[Poutanen et al.(2007)]{poutanen07} Poutanen, J., Lipunova, G., Fabrika, S., Butkevich, A. G., \& Abolmasov, P. 2007, MNRAS, 377, 1187
\bibitem[Roberts(2007)]{roberts07} Roberts, T. P. 2007, Ap\&SS, 311, 203
\bibitem[Saha et al.(2006)]{saha06} Saha, A., Thim, F., Tammann, G. A., Reindl, B., \& Sandage, A. 2006, ApJS, 165, 108
\bibitem[Shakura \& Sunyaev(1973)]{ss73} Shakura, N. I., \& Sunyaev, R. A. 1973, A\&A, 24, 337
\bibitem[Shimura \& Takahara(1995)]{shimura95} Shimura, T. \& Takahara, F. 1995, ApJ, 445, 780
\bibitem[Shrader \& Titarchuk(2003)]{shrader03} Shrader, C. R., \& Titarchuk, L. 2003, ApJ, 598, 168
\bibitem[Sobczak et al.(1999)]{sobczak99} Sobczak, G. J., McClintock, J. E., Remillard, R. A., Bailyn, C. D., \& Orosz, J. A. 1999, ApJ, 520, 776
\bibitem[Sobczak et al.(2000)]{sobczak00} Sobczak, G. J., McClintock, J. E., Remillard, R. A., Cui, W., Levine, A. M., Morgan, E. H., Orosz, J. A., \& Bailyn, C. D. 2000, ApJ, 531, 537
\bibitem[Soria(2007)]{soria07} Soria, R. 2007, Ap\&SS, 311, 213
\bibitem[Soria(2011)]{soria11} Soria, R. 2011, AN, 332, 330
\bibitem[Soria et al.(2012)]{soria12} Soria, R., Kuntz, K. D., 
Winkler, P. F., Blair, W. P., Long, K. S., Plucinsky, P. P., 
\& Whitmore, B. C. 2012, ApJ, 750, 152
\bibitem[Soria et al.(2014)]{soria14} Soria, R., Long, K. S., Blair, W. P., Godfrey, L., Kuntz, K. D., Lenc, E., Stockdale, C., \& Winkler, P. F. 2014, Science, 343, 1330
\bibitem[Steiner et al.(2010)]{steiner10} Steiner, J. F., McClintock, J. E., Remillard, R. A., Gou, L., Yamada, S., \& Narayan, R. 2010, ApJ, 718, L117
\bibitem[Stobbart et al.(2006)]{stobbart06} Stobbart, A.-M., Roberts, T. P., 
\& Wilms, J. 2006, MNRAS, 368, 397
\bibitem[Sutton et al.(2014)]{sutton14} Sutton, A. D., Done, C., \& Roberts, T. P. 2014, MNRAS, 444, 2415
\bibitem[Sutton et al.(2013)]{sutton13} Sutton, A. D., Roberts, T. P., \& Middleton, M. J. 2013, MNRAS, 435, 1758
\bibitem[Thorne (1974)]{thorne74} Thorne, K. S. 1974, ApJ, 191, 507
\bibitem[Tomsick et al.(2005)]{tomsick05} Tomsick, J. A., Corbel, S., Goldwurm, A., \& Kaaret, P. 2005, ApJ, 630, 413
\bibitem[Vierdayanti et al.(2008)]{vierdayanti08} Vierdayanti, K., Watarai, K., \& Mineshige, S. 2008, PASJ, 60, 653
\bibitem[Watarai et al.(2000)]{watarai00} Watarai, K., Fukue, J., \& Mineshige, S. 2000, PASJ, 52, 133
\bibitem[Watarai \& Mineshige(2003)]{watarai03} Watarai, K., \& Mineshige, S. 2003, ApJ, 596, 421
\bibitem[Watarai et al.(2001)]{watarai01} Watarai, K., Mizuno, T., \& Mineshige, S. 2001, ApJ, 549, L77
\end{thebibliography}
\end{document}